# Molecular Dynamics Simulation of the Capillary Leveling of a Glass-Forming Liquid

Ioannis Tanis[1*], Kostas Karatasos[2,3], Thomas Salez[4,5]

[1]Laboratoire de Physico-Chimie Théorique, UMR CNRS Gulliver 7083,
ESPCI Paris, PSL Research University, 75005 Paris, France

[2]Laboratory of Physical Chemistry, Department of Chemical Engineering, Aristotle
University of Thessaloniki, 54124, Thessaloniki

[3] Institute of Electronic Structure and Laser, Foundation for Research and Technology
- Hellas, P. O. Box 1527, 711 10 Heraklion Crete, Greece

[4]Univ. Bordeaux, CNRS, LOMA, UMR 5798, F-33405, Talence, France

[5]Global Station for Soft Matter, Global Institution for Collaborative Research and
Education,
Hokkaido University, Sapporo, Hokkaido 060-0808, Japan

* Corresponding author e-mail address: tanis.ioannis@gmail.com

**Abstract.** Motivated by recent experimental studies probing i) the existence of a mobile layer at the free surface of glasses, and ii) the capillary leveling of polymer nanofilms, we study the evolution of square-wave patterns at the free surface of a generic glass-forming binary Lennard-Jones mixture over a wide temperature range, by means of molecular dynamics simulations. The pattern's amplitude is monitored and the associated decay rate is extracted. The evolution of the latter as a function of temperature exhibits a crossover between two distinct behaviours, over a temperature range typically bounded by the glass-transition temperature and the mode-coupling critical temperature. Layer-resolved analysis of the film particles' mean-squared displacements further shows that diffusion at the surface is considerably faster than in the bulk, below the glass-transition temperature. The diffusion coefficient of the surface particles is larger than its bulk counterpart by a factor that reaches $10^5$ at the lowest temperature studied. This factor decreases upon heating, in agreement with recent experimental studies.

## I. Introduction

Thin films and their free interfaces are used in a variety of applications such as catalysis[1], fast crystal growth[2, 3] and the formation of low-energy glasses[4, 5]. Apart from their applications in industry and technology, molecular mobility and relaxation at surfaces and interfaces have been recently a subject of growing interest for fundamental research. To this end, several experimental approaches have been used to probe the surface evolution and mobility of thin films[6-10]. More specifically, an efficient way to gain insight into the surface mass transport and the associated dynamics, is to deposit nanoparticles on the film surface and subsequently to monitor the host material's response[9,11]. Another powerful and versatile technique widely utilized for the determination of the surface-diffusion coefficient $D_s$, characterizing the in-plane translation of molecules at the film surface, is the capillary-driven leveling of surface-gratings[6, 9, 12-14]. Several studies suggest that the mechanism governing the relaxation of these gratings is temperature and material dependent[12, 15]. More specifically, recent works on glass-forming molecular liquids reported a transition from bulk viscous flow at temperatures higher than the glass-transition temperature $T_g$, to surface diffusion below $T_g$ [13, 16]. Analogous studies in polymeric liquids revealed that bulk viscous flow was the only mechanism governing the surface decay, even at the highest viscosities studied [14, 17, 18]. According to Mullin's pioneering work[19], the different mechanisms that could flatten the surface pattern (surface and bulk diffusions, evaporation-condensation, bulk viscous flow) can be distinguished by determining the decay rate dependence on spatial frequency. This method has been applied in crystalline metals[20], amorphous silica[21] and molecular glasses[22].

Measurements of diffusion coefficients at the surface report values considerably higher than the respective bulk coefficients at the glass-transition temperature[12,13]. Furthermore, large variations in surface diffusion coefficients among different molecular glass formers bearing a similar bulk mobility were linked to the strength of the intermolecular forces[23].

Further insight into glass-forming systems has been provided by numerical simulations. Several studies have addressed the factors that govern dynamical heterogeneity and spatial correlations in the bulk[24-30]. Furthermore, other works addressing systems under confinement have managed to elucidate the effects of confinement by rough or smooth walls on the liquid dynamics and to extract associated length-scales[31-35]. Extensive simulation studies have also been conducted on ultrastable vapor-deposited glasses [4, 36, 37]. However, to our knowledge, a small number of simulation studies probing the surface mobility in glass formers have been reported so far.

Surface-diffusion-mediated decay of two-dimensional nanostructures has been recently studied by a combined analytical and kinetic Monte-Carlo approach[38], whereas Kayhani *et al*. performed molecular dynamics (MD) simulations to examine the coalescence of platinum nanoclusters[39]. Concerning simulation work on the surface mobility of molecular glass formers, Hoang *et al*. performed MD studies on freestanding monatomic glass-forming liquids[40]. The study of Akbari *et al*. addressed the effects of the substrate in kinetic and thermodynamic properties of a binary Lennard-Jones (LJ) liquid[41], whereas Malshe and co-workers extracted surface-diffusion coefficients by the aid of the grating-decay approach[42] in free standing films of the aforementioned liquid type. Results were consistent with data obtained from the mean-squared displacements of the liquid particles. In a very recent study, Kuan and co-workers examined single-particle dynamics in the surface and in the bulk of supported films prepared by vapour deposition. Both bulk and surface particle displacements showed evidence of heterogeneous dynamics[43].

The aim of the present work is to provide additional insight into the surface mobility of a glass-forming liquid by means of molecular dynamics simulations. Similarly to a recent work where we probed the viscoelastic behaviour of nonentangled polymer films above the glass-transition temperature, through the capillary leveling of a square-wave surface pattern[44], we examine here the evolution of a square-wave surface pattern atop a generic binary LJ mixture supported by an attractive substrate. This new study is carried out over a wide temperature range, sampling both the glassy and liquid states of the film. The associated decay rates are extracted whereas surface and bulk mobilities are examined. Further insight into the surface and bulk mobilities is gained by examining the self part of the van Hove function, as well as via the layer-resolved analysis of the liquid particles' mean-squared displacements.

## II. Computational Details

We examine a binary mixture of 80% A and 20% B particles that interact via a LJ potential $V_{\alpha\beta} = 4\epsilon_{\alpha\beta} \left[ \left( \sigma_{\alpha\beta}/r \right)^{12} - \left( \sigma_{\alpha\beta}/r \right)^{6} \right]$, $(\alpha, \beta \in \{A, B\})$. The Van der Waals parameters were chosen as $\epsilon_{AA} = 1.0, \sigma_{AA} = 1.0, \epsilon_{AB} = 1.5, \sigma_{AB} = 0.8, \epsilon_{BB} =$ $0.5, \sigma_{BB} = 0.88$. The potential was truncated at a cut-off radius $r_c = 2.5\sigma_{AB}$. The numbers of particles of type A and B are 2938 and 734, respectively. The binary mixture interacting with this parameter set corresponds to the Kob-Andersen liquid model which is not prone to crystallization or to phase separation; it is extensively utilized in studies of glass-forming liquids[45, 46]. The liquid film is supported by a strongly attractive substrate whose atoms (henceforth designated by the type S) are located on the sites of an hexagonal lattice and are kept fixed to their lattice positions. The substrate particles' radius is $\sigma_S = 2.41$ and they attract A and B atoms with the following energy magnitudes[47]: $\epsilon_{AS} = 4.425$ and $\epsilon_{BS} = 3.129$. The length of the cubic box is $16.56\sigma_{AA}$ in the lateral directions $x$ and $y$, and $441.17\sigma_{AA}$ in the $z$ dimension. The value of the latter is chosen large enough to avoid interaction of the particles with periodic images of the substrate, and ensure a system pressure $p = 0$. All quantities are reported in LJ units, *i.e.* length is measured in units of $\sigma_{AA}$, energy in units of $\epsilon_{AA}$, time in units of $\tau = (m\sigma_{AA}^2/(48\epsilon_{AA}))^{1/2}$ and temperature in units of $\epsilon_{AA}/k_B$ where $k_B$ stands for the Boltzmann constant. For the case of argon, these units correspond to a length of 3.4 Å, an energy of $120Kk_B$ and a time of $3\cdot10^{-13}$ s. All simulations are performed in the canonical (*NVT*) ensemble using the DL_POLY code[48]. Temperature is maintained constant through the use of the Nosé-Hoover thermostat. Simulations are conducted at the temperature range $T = 0.20$-$0.525$, and the length of each run is $3\cdot10^5$.

The initial configuration of the glassy film was obtained by the aid of the Aten program[49]. To create the 3-dimensional pattern at the free surface of the film, a flat film was initially equilibrated in the canonical ensemble at a temperature of $T = 0.3$. Afterwards, particles were removed from both edges of the free surface in order to obtain the 3D pattern, henceforth designated as the 'top layer' (see Fig.1). The lateral dimensions of the latter were $L_x = L_y = 4.41$, whereas its thickness and initial height (counted from the substrate) were respectively $d = 4.36$ and $h_0 = 15.67$.

To examine the behaviour in the bulk, a configuration comprising 5076 particles was constructed at an elevated temperature, $T = 0.80$, and at a density of $\rho = 1.2$. After removing close contacts, the system was cooled to $T = 0.30$ by conducting runs in the isothermal-isobaric ensemble (*NPT*) at pressure $p = 0$. The temperature step between the isothermal runs ranged from $dT = 0.1$ for the highest temperatures to $dT = 0.025$ and $dT = 0.0125$ at temperatures around the bulk glass-transition temperature $T_g$. The duration of each isothermal run ranged from $1.6 \cdot 10^4$ (at $T > 0.5$) to $3 \cdot 10^5$ (at $T \leq 0.5$). Defining the cooling rate as the difference between the starting temperature and the final temperature divided by the time of the quench, the cooling rate was $0.75 \cdot 10^{-7}$. This amounts to half the cooling rate used in the original model of Kob and Andersen and, although it remains quite fast, it is slower than the fastest cooling rate in experiments[45]. The glass-transition temperature of the binary mixture was determined by extrapolating and intersecting the affine dependencies of the low- and high-temperature specific-volume-vs-temperature curves. This yielded a value of $T_g = 0.392$ which is below the reported mode-coupling critical temperature for this system, $T_c \approx 0.435$[50].

Figure 1 displays the initial configuration of the samples, as well as snapshots from a run at $T = 0.3$ at times $t = 20000$ and $233333$. All snapshots were generated by the aid of the VMD software[51].The particles that evaporated at elevated temperatures were disregarded from post-analysis.

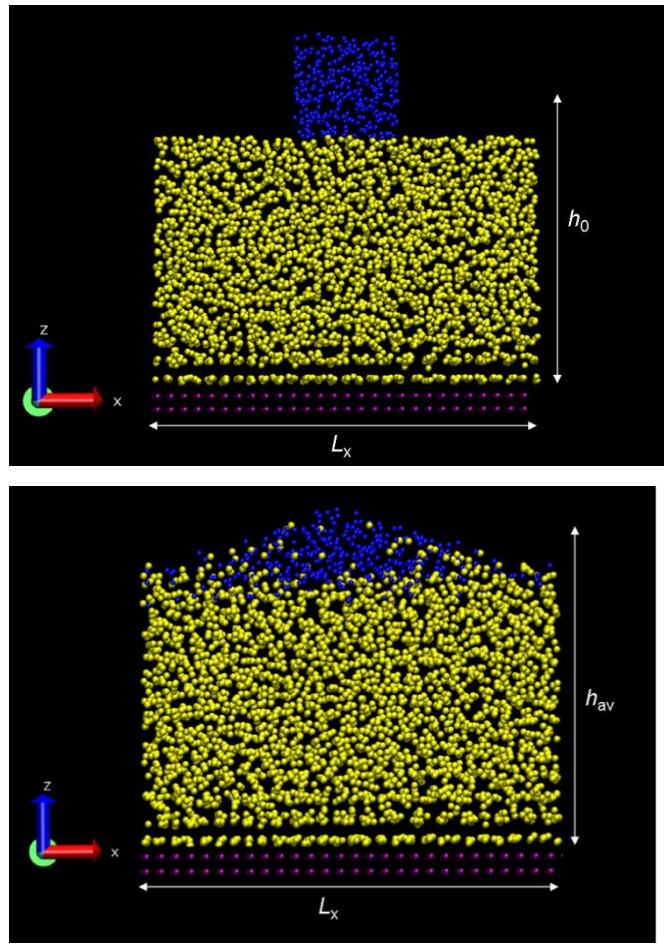

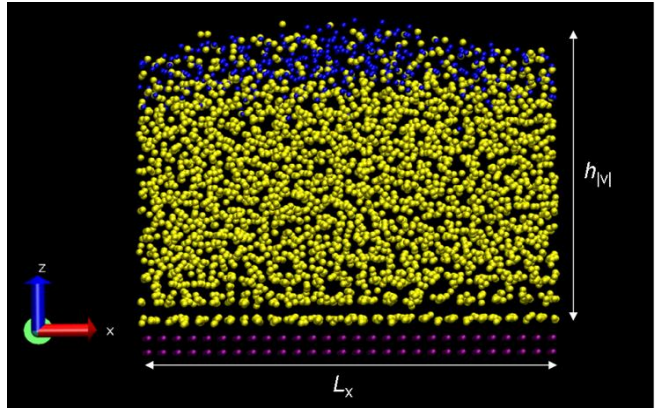

Figure 1: Top: Initial configuration of the sample bearing a total number of 3672 particles. Periodic boundary conditions are applied in all directions. $L_x$ stands for the $x$-dimension of the simulation box and $h_0$ is the maximum vertical height (colour code : blue = top layer, yellow = bottom layer, purple = substrate). Middle: Snapshot of the film at $t = 20000$, where $h_{av}$ is the average height value. Bottom: Snapshot of the film at $t = 233333$; $h_{lvl}$ the average height of the fully leveled film.

### III. Results

To investigate the mobility of the liquid's molecules, we examine the self-part of the van Hove space-time correlation function defined as :

$$G_s(x,t) = \frac{1}{N}\sum_{i=1}^{N}\langle \delta\big(x + x_i(0) - x_i(t)\big)\rangle, \qquad (1)$$

with $x_i$ representing the position of particle $i$ along any of the $x$, $y$ and $z$ directions, $N$ being the number of particles that constitute the top layer, $\langle ... \rangle$ being the ensemble average over realizations and time origins and $\delta$ is the Dirac's distribution. In order to account for the natural anisotropy of the system induced by the interfaces, the van Hove function was evaluated in each direction separately. In Fig.2, the $x$ dependencies of $4\pi x^2 G_s(x,t)$ for the top layer particles and two times $t$, are displayed. A practically indistinguishable behaviour (not shown for clarity reasons) was observed in the $y$ direction, as well. Also shown in the figure, are the corresponding curves for the bulk sample. The van Hove functions were evaluated at the temperatures $T = 0.30$ and $T = 0.35$, which lie below the bulk $T_g$ as well as at a temperature $T = 0.375$ near the bulk $T_g$.

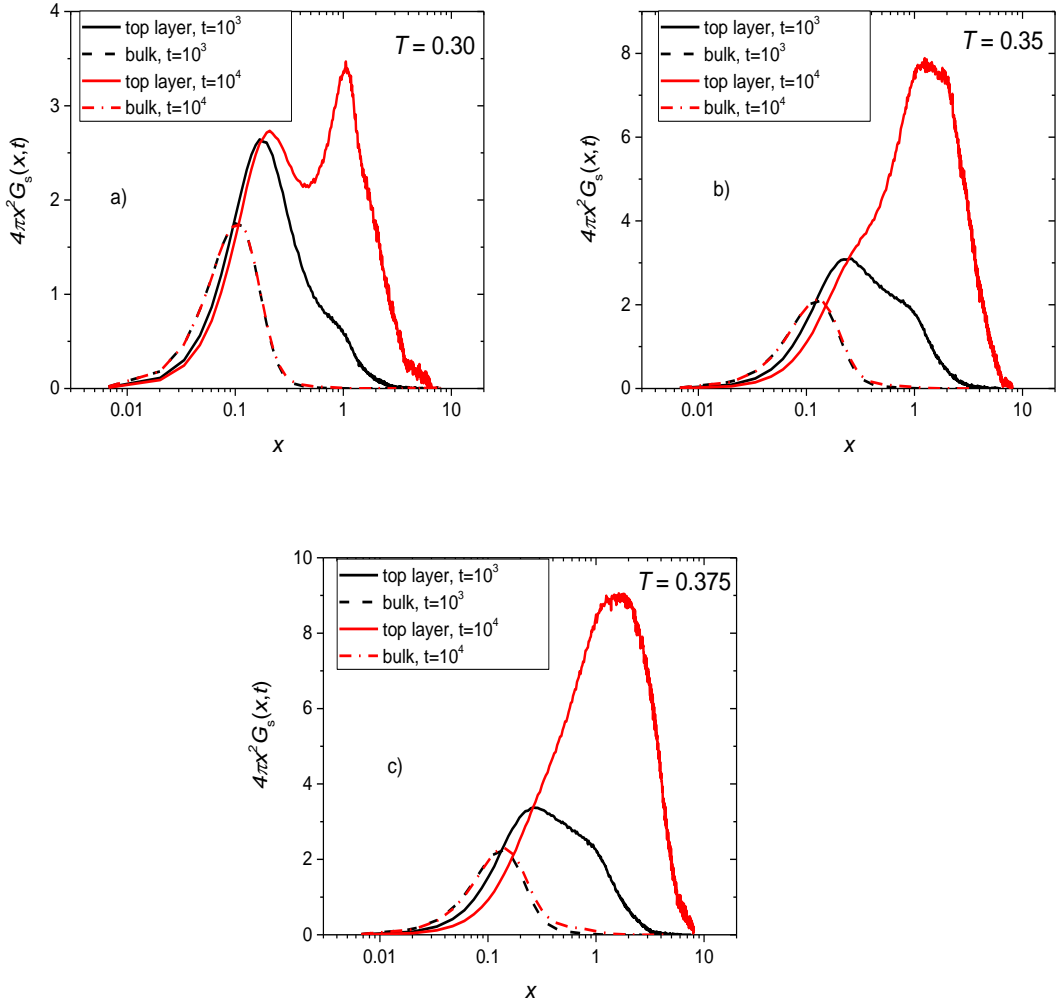

Figure 2: Self part of the van Hove function (eq.1) for the top layer particles, as a function of the molecular displacement in the $x$ direction, evaluated at a) $T = 0.30$, b) $T = 0.35$ and c) $T = 0.375$. The different line colours correspond to different times and the references for a bulk sample are provided for comparison.

A visual inspection of Fig.2 reveals that the top layer particles are significantly more mobile than those corresponding to the bulk. Focusing on the top left panel, we observe that in the bulk, and at low temperature, the distribution shows a vanishing probability for $x$-displacement larger than the near-neighbour distance ($x \sim 1$). As expected for a glassy behaviour, this implies that the particles have not managed to escape from the cages in which they were trapped initially. In contrast, the distribution for the top layer particles exhibits non-negligible values up to $x \sim 4$, thus providing evidence for an enhanced mobility at the free interface of the film. As expected, the difference between surface and bulk mobilities grows with time. Bulk mobility remains practically unchanged at the timescales examined, since the system at the temperatures shown in Fig.2 resides below the estimated glass-transition temperature. Focusing on the top layer particles' distribution at early times, one can discern, apart from the peak at

$x \sim 0.1$, the existence of a "shoulder" at distance $x \sim 1$. The distribution at later times further confirms the existence of two different populations, as reflected in the peaks at $x \sim 0.2$ and at $x \sim 1.1$. The appearance of these two peaks reflects a clear distinction in mobility for the surface and internal particles belonging to the top-layer and lies in accordance with the findings of Malshe et al.[42]. On the other hand, the distribution also shows that the peak belonging to the more mobile particles gains significantly in intensity as time lapses. This is consistent with the fact that an increasing fraction of the top-layer particles gets closer to the free surface with time (as demonstrated by the snapshots in Fig.1, blue particles).

The previous two-peak fine-structure separation of relaxation processes is no longer detected at late times for $T > 0.30$, thus indicating a more homogenous mobility at higher temperature. Nevertheless, the "shoulder" observed in panels b and c at $x \sim 1$ at early times, is maintained and gains significantly in intensity at higher temperatures. Assuming that the mechanism of diffusion does not change appreciably at the sub-$T_g$ temperatures examined, it is reasonable to surmise that this behavior arises from the same origin as in the lowest temperature case, *i.e.* the increase with time of the number of top-layer particles gaining access to the free surface. In sharp contrast, there is no sign of enhanced mobility in the bulk samples as temperature rises. Only at $T = 0.375$, one can discern a marginal increase in the mobility of the bulk particles at long times, as reflected by the tail of the van Hove distribution. Since the bimodal distribution at low temperatures which is indicative for the dynamic heterogeneity in the motion of the top layer particles, is no longer detected at elevated temperatures, it may be considered as a signature for the transition for these particles to the glassy state.

To examine the global consequences of the enhanced surface mobility on the film leveling, we have monitored the average height $h_{av}(t)$ of the top layer particles. The initial value of $h_{av}$ (counted from the substrate) was $h_{av0} = 13.5$. An example of the time evolution of the rescaled amplitude, $\frac{h_{av}(t) - h_{|v|}}{h_{av0} - h_{|v|}}$, for a temperature $T = 0.45$, is displayed in Fig.3a, where $h_{|v|} = 11.31$ is the average height of the top layer particles

after the film leveling. It should be noted here that $h_{|v|}$ is smaller than the value predicted from the volume conservation. This is due to the fact that, in this work, it is the average height of the square pattern that is monitored rather than its maximum height. On these grounds, $h_{|v|}$ does not correspond the value of $h_{av}(t)$ at infinite time and, consequently, it is not an equilibrium value. We stress that the total number of evaporated particles did not exceed 0.35% of the total number of particles constituting the film, for all simulated systems. As stated earlier, the evaporated particles were excluded from all analysis calculations. As it can be inferred from Fig.3a, the temporal decay is exponential. Figure 3b displays the associated decay constant as a function of temperature. We note that the minimum temperature at which full leveling of the square pattern was observed within the total simulation time, was $T = 0.30$.

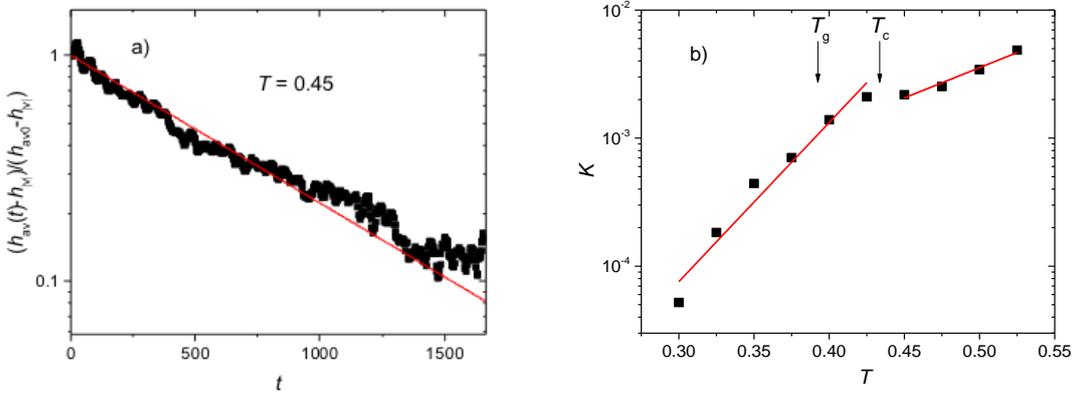

Figure 3: a) Time evolution of the rescaled pattern amplitude at $T = 0.45$. The red line is an affine fit. b) Decay constant $K$ extracted from the affine fits as a function of temperature. Error bars in $K$ are on the order of $10^{-6}$.

Examining Fig.3b, it appears that in between $T_g = 0.392$ and $T_c = 0.435$, a crossover takes place. At temperatures above $T_c$, a weak slope characteristic of a liquid-like behaviour is observed, while below $T_g$ a stronger dependence corresponding to a glassy-like behaviour is noticed. This observation is consistent with the fact that the onset of supercooling effects is generally observed near $T_c$. Interestingly, recent experimental studies report different rates for the decay of patterned surfaces in supported films above and below the glass-transition temperature[6, 9, 11, 13, 23]. The flattening of profile patterns at the free surface of films below $T_g$ has been ascribed to surface diffusion as suggested by Mullin's theory[19]. It must be noted that, since the mobile layer can still be considered at equilibrium within a certain temperature range below $T_g$, and thus the Stokes-Einstein relation is valid there, one can also describe the surface mobility as a surface flow which leads to an identical partial differential equation for the profile's evolution[9]. Surface diffusion levels a sinusoidal surface pattern at a decay rate proportional to $q^4$, where $q$ is the angular-wavenumber. Other mechanisms, such as bulk viscous flow and evaporation-condensation exhibit weaker $q$ dependences (except for the limiting lubrication case[9], that accidently shows the same $q^4$ dependence for a bulk viscous flow above $T_g$). We have not investigated the angular- wavenumber dependence of the decay rate here but, instead, we have carried out a layer-resolved analysis of the particles' $z$-dependent mean-squared displacements (MSDs) as defined by:

$$\Delta r^2(z,t) = \left\langle \frac{1}{n_t} \sum_i |r_i(t) - r_i(0)|^2 \prod_{t'=0}^{t} \delta[z - z_i(t')] \right\rangle, \qquad (2)$$

where $r_i(t)$ is the position of particle $i$ at time $t$ and $\langle ... \rangle$ is an ensemble average. This definition takes into account only the $n_t$ particles which are at all times $t' < t$ within the slab centred at $z$ and bearing a width of $\Delta z = 1.5$. The results for the runs carried out at $T = 0.35$ are presented in Fig.4. For comparison purposes, the results for runs carried out above $T_g$, at $T = 0.46$, are also shown. We note here that the simulation runs at $T = 0.46$ were significantly shorter than the runs below $T_g$ due to the evaporation of the surface particles.

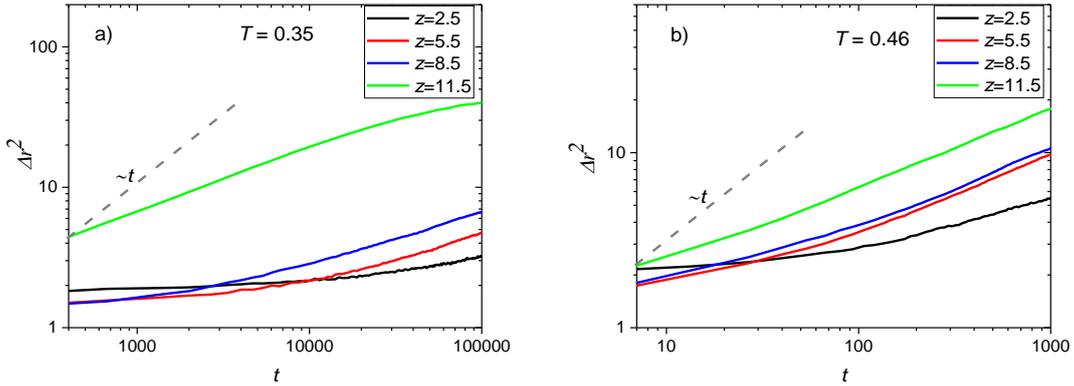

Figure 4: Layer-resolved mean-squared displacements (see eq.2) as a function of time, for various distances $z$ from the substrate as indicated, and at temperatures a) $T = 0.35$ and b) $T = 0.46$. The dashed line corresponds to the scaling law $\Delta r^2 \sim t$.

Focusing on the left panel of Fig.4, it appears that, at $T = 0.35$, the MSD of the particles located in the outermost film layer near the film surface is about 5 times larger than the one for the particles in the second outermost film layer. There is no sign of major propagation of the enhanced surface mobility into the bulk. This observation corroborates a scenario in which the relaxation of the square pattern is governed by surface diffusion below the glass transition. A similar behaviour was observed at all sampled temperatures below $T_g$. Furthermore, a visual inspection of the MSDs of the innermost particles (black lines in Fig.4), demonstrates enhanced dynamics near the solid substrate at short times. This behaviour lies in accordance with simulation studies of polymers near solid interfaces[52,53,54].

Finally, the apparent total diffusion coefficients, $D_s = \lim_{t\to\infty} \frac{1}{6t} \Delta r^2$, of the top layer particles were extracted from the MSD curves at different temperatures, and compared with the corresponding bulk values, $D_b$ in Fig.5. We should note that, although the diffusion of the top layer particles is anisotropic, we have calculated the total diffusion coefficient in order to be consistent with the calculations in the bulk case.

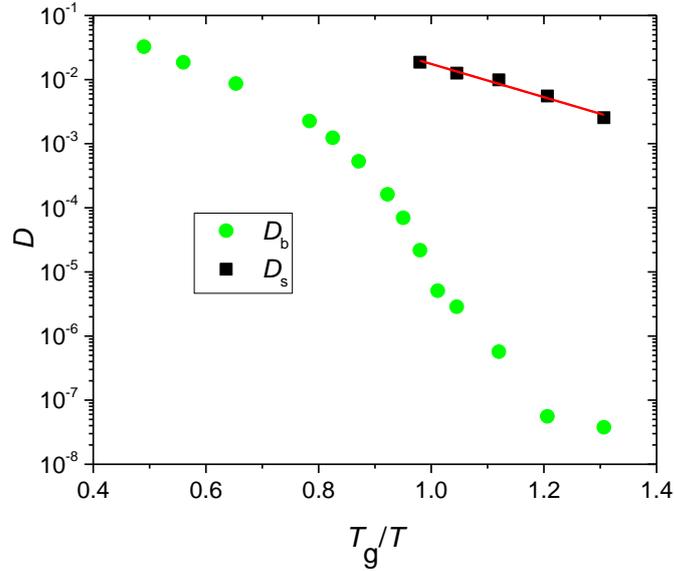

Figure 5: Apparent diffusion coefficients of the top layer ($D_s$) and bulk ($D_b$) particles versus inverse temperature. At temperatures $T \leq 0.45$, data points correspond to estimations in the non-Fickian regime. Temperature is scaled by the bulk $T_g$. The solid line is an affine fit in log-linear representation.

As can be discerned, the diffusion coefficient of the top layer particles seems to exhibit an Arrhenius-type temperature dependence, which is reminiscent of previous observations for oligomer glasses[6, 9]. Moreover, diffusion at the surface is faster than in the bulk by a factor that ranges from $10^2$ to $10^5$ at the lowest temperature studied. This ratio drastically rises with cooling as it has been observed in both experimental and theoretical studies[6, 9, 13, 42].

**IV. Conclusion**

Molecular dynamics simulations have been conducted in order to examine the decay process of a periodic square-wave pattern at the free surface of a binary Lennard-Jones film. Contrary to previous simulation studies of the capillary leveling of thin films, the evolution of the height was examined over a wide temperature range sampling both the liquid and glassy states of the film. Different mechanisms appeared to control the pattern leveling in those two states. By the aid of layer-resolved analysis of the particles' mean-squared displacements, we showed that diffusion is much more efficient in the layer close to the free surface than elsewhere in the bulk. Specifically, at the lowest temperature studied, the diffusion coefficient of the surface particles is found to be $10^5$ times larger than the bulk counterpart and this difference decreases upon heating.

**Acknowledgements**

The authors gratefully acknowledge financial support from ANR WAFPI and ANR FSCF, as well as the Global Station for Soft Matter, a project of Global Institution for Collaborative Research and Education at Hokkaido University. Hendrik Meyer is kindly thanked for providing the layer-resolved analysis code. The authors also thank Anthony Maggs, Elie Raphaël, Kari Dalnoki-Veress, James Forrest, Joshua McGraw, Oliver Bäumchen and Jörg Baschnagel for stimulating discussions. K.K. would like to thank FO.R.TH.-IESL for the warm hospitality during his stay there.

**References**

(1)     Ulbricht, M. Advanced functional polymer membranes. *Polymer* **2006,** *47*, 2217-2262.

(2)     Hasebe, M.; Musumeci, D.; Powell, C. T.; Cai, T.; Gunn, E.; Zhu, L.; Yu, L. Fast surface crystal growth on molecular glasses and its termination by the onset of fluidity. *J. Phys. Chem. B* **2014,** *118*, 7638-7646.

(3)     Sun, Y.; Zhu, L.; Kearns, K. L.; Ediger, M. D.; Yu, L. Glasses crystallize rapidly at free surfaces by growing crystals upward. *Proc. Natl. Acad. Sci. U.S.A.* **2011,** *108*, 5990-5995.

(4)     Ediger, M. D. Vapor-deposited glasses provide clearer view of two-level systems. *Proc. Natl. Acad. Sci. U.S.A.* **2014,** *111*, 11232-11233.

(5)     Mangalara, J. H.; Marvin, M. D.; Simmons, D. S. Three-layer model for the emergence of ultrastable glasses from the surfaces of supercooled liquids. *J. Phys. Chem. B* **2016,** *120*, 4861-4865.

(6)     Yang, Z.; Fujii, Y.; Lee, F. K.; Lam, C.-H.; Tsui, O. K. C. Glass transition dynamics and surface layer mobility in unentangled polystyrene films. *Science* **2010,** *328*, 1676-1679.

(7)     Yang, Z.; Clough, A.; Lam, C.-H.; Tsui, O. K. C. Glass transition dynamics and surface mobility of entangled polystyrene films at equilibrium. *Macromolecules* **2011,** *44*, 8294-8300.

(8)     Mapes, M. K.; Swallen, S. F.; Ediger, M. D. Self-diffusion of supercooled o-terphenyl near the glass transition temperature. *J. Phys. Chem. B* **2006,** *110*, 507-511.

(9)     Chai, Y.; Salez, T.; McGraw, J. D.; Benzaquen, M.; Dalnoki-Veress, K.; Raphaël, E.; Forrest, J. A. A direct quantitative measure of surface mobility in a glassy polymer. *Science* **2014,** *343*, 994-999.

(10)    Teisseire, J.; Revaux, A.; Foresti, M.; Barthel, E. Confinement and flow dynamics in thin polymer films for nanoimprint lithography. *Appl. Phys. Lett.* **2011,** *98*, 013106.

(11)    Fakhraai, Z.; Forrest, J. A. Measuring the surface dynamics of glassy polymers. *Science* **2008,** *319*, 600.

(12)     Zhang, W.; Yu, L. Surface diffusion of polymer glasses. *Macromolecules* **2016,** *49*, 731-735.

(13)    Brian, C. W.; Yu, L. Surface self-diffusion of organic glasses. *J. Phys. Chem. A* **2013,** *117*, 13303-13309.

(14)    Hamdorf, M.; Johannsmann, D. Surface-rheological measurements on glass forming polymers based on the surface tension driven decay of imprinted corrugation gratings. *J. Chem. Phys.* **2000,** *112*, 4262-4270.

(15)    Zhang, W.; Brian, C. W.; Yu, L. Fast surface diffusion of amorphous o-terphenyl and its competition with viscous flow in surface evolution. *J. Phys. Chem. B* **2015,** *119*, 5071-5078.

(16)    Ruan, S.; Musumeci, D.; Zhang, W.; Gujral, A.; Ediger, M. D.; Yu, L. Surface transport mechanisms in molecular glasses probed by the exposure of nano-particles. *J. Chem. Phys.* **2017,** *146*, 203324.

(17)    Kim, H.; Rühm, A.; Lurio, L. B.; Basu, J. K.; Lal, J.; Lumma, D.; Mochrie, S. G. J.; Sinha, S. K. Surface dynamics of polymer films. *Phys. Rev. Lett.* **2003,** *90*, 068302.

(18)    Wang, L.; Ellison, A. J. G.; Ast, D. G. Investigation of surface mass transport in Al–Si–Ca–oxide glasses via the thermal induced decay of submicron surface gratings. *J. Appl. Phys.* **2007,** *101*, 023530.

(19)    Mullins, W. W. Flattening of a nearly plane solid surface due to capillarity. *J. Appl. Phys.* **1959,** *30*, 77-83.

(20)    Bonzel, H. P.; Latta, E. E. Surface self-diffusion on Ni(110): Temperature dependence and directional anisotropy. *Surf. Sci.* **1978,** *76*, 275-295.

(21)    Keeffe, M. E.; Umbach, C. C.; Blakely, J. M. Surface self-diffusion on Si from the evolution of periodic atomic step arrays. *J. Phys. Chem. Solids* **1994,** *55*, 965-973.

(22)    Zhu, L.; Brian, C. W.; Swallen, S. F.; Straus, P. T.; Ediger, M. D.; Yu, L. Surface self-diffusion of an organic glass. *Phys. Rev. Lett.* **2011,** *106*, 256103.

(23)    Chen, Y.; Zhang, W.; Yu, L. Hydrogen bonding slows down surface diffusion of molecular glasses. *J. Phys. Chem. B* **2016,** *120*, 8007-8015.

(24)    Donati, C.; Glotzer, S. C.; Poole, P. H.; Kob, W.; Plimpton, S. J. Spatial correlations of mobility and immobility in a glass-forming Lennard-Jones liquid. *Phys. Rev. E* **1999,** *60*, 3107-3119.

(25)    Kob, W.; Donati, C.; Plimpton, S. J.; Poole, P. H.; Glotzer, S. C. Dynamical heterogeneities in a supercooled Lennard-Jones liquid. *Phys. Rev. Lett.* **1997,** *79*, 2827-2830.

(26)    Andersen, H. C. Molecular dynamics studies of heterogeneous dynamics and dynamic crossover in supercooled atomic liquids. *Proc. Natl. Acad. Sci. U.S.A.* **2005,** *102*, 6686-6691.

(27)    Kob, W. Computer simulations of supercooled liquids and glasses. *J. Phys. Condens. Matter* **1999,** *11*, R85.

(28)    Murthy, S. S. N. Molecular dynamics in supercooled liquids: A study of the relaxation in binary solutions. *J. Mol. Liq.* **1992,** *51*, 197-217.

(29)    Pakula, T. Collective dynamics in simple supercooled and polymer liquids. *J. Mol. Liq.* **2000,** *86*, 109-121.

(30)     Vogel, M.; Medick, P.; Rossler, E. Slow molecular dynamics in binary organic glass formers. *J. Mol. Liq.* **2000,** *86*, 103-108.

(31)     Hocky, G. M.; Berthier, L.; Kob, W.; Reichman, D. R. Crossovers in the dynamics of supercooled liquids probed by an amorphous wall. *Phys. Rev. E* **2014,** *89*, 052311.

(32)     Scheidler, P.; Kob, W.; Binder, K. Cooperative motion and growing length scales in supercooled confined liquids. *EPL Europhys. Lett.* **2002,** *59*, 701.

(33)     Varnik, F.; Scheidler, P.; Baschnagel, J.; Kob, W.; Binder, K. Molecular dynamics simulation of confined glass forming liquids. *Mater. Res. Soc. Symp. Proc.* **2000,** *651*, T3.1.1.

(34)     Scheidler, P.; Kob, W.; Binder, K. The relaxation dynamics of a simple glass former confined in a pore. *EPL Europhys. Lett.* **2000,** *52*, 277.

(35)     Scheidler, P.; Kob, W.; Binder, K. The relaxation dynamics of a supercooled liquid confined by rough walls. *J. Phys. Chem. B* **2004,** *108*, 6673-6686.

(36)     Singh, S.; Ediger, M. D.; de Pablo, J. J. Ultrastable glasses from in silico vapour deposition. *Nat. Mater.* **2013,** *12*, 139.

(37)     Lin, P.-H.; Lyubimov, I.; Yu, L.; Ediger, M. D.; de Pablo, J. J. Molecular modeling of vapor-deposited polymer glasses. *J. Chem. Phys.* **2014,** *140*, 204504.

(38)     Marcos, F. C.; Ezequiel, V. A. Continuous and discrete modeling of the decay of two-dimensional nanostructures. *J. Phys. Condens. Matter* **2009,** *21*, 263001.

(39)     Kayhani, K.; Mirabbaszadeh, K.; Nayebi, P.; Mohandesi, A. Surface effect on the coalescence of Pt clusters: A molecular dynamics study. *Applied Surf. Sci.* **2010,** *256*, 6982-6985.

(40)     Hoang, V. V.; Dong, T. Q. Free surface effects on thermodynamics and glass formation in simple monatomic supercooled liquids. *Phys. Rev. B* **2011,** *84*, 174204.

(41)     Haji-Akbari, A.; Debenedetti, P. G. The effect of substrate on thermodynamic and kinetic anisotropies in atomic thin films. *J. Chem. Phys.* **2014,** *141*, 024506.

(42)     Malshe, R.; Ediger, M. D.; Yu, L.; de Pablo, J. J. Evolution of glassy gratings with variable aspect ratios under surface diffusion. *J. Chem. Phys.* **2011,** *134*, 194704.

(43)     Kuon, N.; Flenner, E.; Szamel, G. Comparison of single particle dynamics at the center and on the surface of equilibrium glassy films. *J. Chem. Phys.* **2018,** *149*, 074501.

(44)    Tanis, I.; Meyer, H.; Salez, T.; Raphaël, E.; Maggs, A. C.; Baschnagel, J. Molecular dynamics simulation of the capillary leveling of viscoelastic polymer films. *J. Chem. Phys.* **2017**, *146*, 203327.

(45)    Kob, W.; Andersen, H. C. Testing mode-coupling theory for a supercooled binary Lennard-Jones mixture. *Transport Theor. Stat. Phys.* **1995**, *24*, 1179-1198.

(46)    Kob, W.; Barrat, J.-L. Aging effects in a Lennard-Jones glass. *Phys. Rev. Lett.* **1997**, *78*, 4581-4584.

(47)    Li, R.; Wang, L.; Yue, Q.; Li, H.; Xu, S.; Liu, J. Insights into the adsorption of oxygen and water on low-index Pt surfaces by molecular dynamics simulations. *New J. Chem.* **2014**, *38*, 683-692.

(48)    Smith, W.; Todorov, I. T. A short description of DL_POLY. *Mol. Simul.* **2006**, *32*, 935-943.

(49)    Youngs, T. G. A. Aten—An application for the creation, editing, and visualization of coordinates for glasses, liquids, crystals, and molecules. *J. Comput. Chem.* **2009**, *31*, 639-648.

(50)    Kob, W.; Andersen, H. C. Scaling behavior in the beta-relaxation regime of a supercooled Lennard-Jones mixture. *Phys. Rev. Lett.* **1994**, *73*, 1376-1379.

(51)    Humphrey, W.; Dalke, A.; Schulten, K. VMD: Visual molecular dynamics, *J. Mol. Graph.* **1996,** *14*, 33-38.

(52)    Daoulas, K. C.; Harmandaris, V. A.; Mavrantzas, V. G. Molecular dynamics simulation of a polymer melt/solid interface: local dynamics and chain mobility in a thin film of polyethylene melt adsorbed on graphite. *Macromolecules* **2005**, *38*, 5796–5809.

(53)    Kritikos, G.; Sgouros, A.; Vogiatzis, G. G.; Theodorou, D. N. Molecular dynamics study of polyethylene under extreme confinement. *J. Phys. Conf. Ser.* **2016**, 738, 012012.

(54)    Baschnagel, J.; Binder, K.; Milchev, A. Mobility of polymers near surfaces. In *Polymer Surfaces, Interfaces and Thin Films* **2000**, 1–49.

TOC Graphic

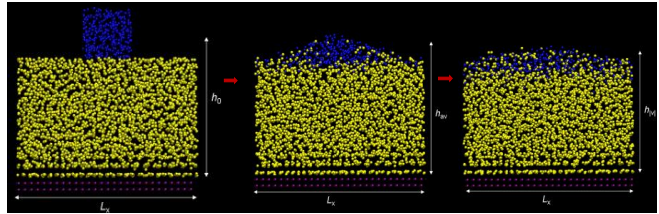